\newcommand{\ben}{\begin{equation}}
\newcommand{\bal}{\begin{align}}
\newcommand{\een}{\end{equation}}
\newcommand{\eal}{\end{align}}
\newcommand{\bea}{\begin{eqnarray}}
\newcommand{\eea}{\end{eqnarray}}
\newcommand{\nn}{\nonumber\\ }

\newcommand{\qq}{\qquad\qquad}
\newcommand{\QQ}{\qquad\qquad\qquad\qquad}
\newcommand{\al}{{\alpha}}

\newcommand{\infinity}{{\infty}}

\newcommand{\cW}{{\cal W}}
\newcommand{\hfb}{{\hfill\break}}
\newcommand{\sub}[1]{_{\stackrel{}{#1}}}
\input amssym.def
 \def\R{{\Bbb R}}  \def\N{{\Bbb N}}
   
\documentclass{article}
\usepackage{graphicx}
\usepackage{epsfig}

\begin{document}

\parskip=4pt
\baselineskip=14pt
%%%%%%%%%%%% Title %%%%%%%%%%

\title{\vskip-1cm The Morse potential and phase-space quantum mechanics}
\author{B. Belchev, M.A. Walton\\\\ {\it Department of
Physics and Astronomy,
University of Lethbridge}\\
{\it Lethbridge, Alberta, Canada\ \  T1K 3M4}\\\\
{\small borislav.belchev@uleth.ca, walton@uleth.ca}\\\\
}

\maketitle
%%%%%%%%%%%%%%%%%%%%%%%%%%%%%%%%%%%%%%%%%%%%%%%%%%%%%%%%%%%%%%%%%%%%%%%%
\begin{abstract}

We consider the time-independent Wigner functions  of phase-space quantum mechanics (a.k.a. deformation quantization)  for a
Morse potential. First, we find them by solving the $\ast$-eigenvalue equations, using a method that can  be applied to
potentials that are polynomial in an exponential.  A Mellin transform converts the $\ast$-eigenvalue equations to difference
equations, and factorized solutions are found directly for all values of the parameters. The symbols of both diagonal and
off-diagonal density operator elements in the energy basis are found this way. The Wigner transforms of the density matrices
built from the known wave functions are then shown to confirm the solutions.

\end{abstract}


\vfill\eject

\section{Introduction}

Phase-space quantum mechanics (see \cite{ZFC,HWW}, e.g.) is also known as deformation quantization (see \cite{BFFLS}). In it,
the density operator is replaced by its symbol, the Wigner function (or distribution) on phase space, and its equations of
motion involve the Moyal $\ast$-product.  Although a completely autonomous formulation of quantum mechanics results, the
$\ast$-eigenvalue equations obtained are difficult to solve. Few closed-form solutions have been found; see \cite{ZFC,CFZ} and
references therein. Here we derive a new, explicit solution of the $\ast$-eigenvalue equations of phase-space quantum mechanics
by treating the Morse potential.

The Morse potential is used in various physical applications. Our motivation, however, came from the deformation quantization
of contact interactions. In particular, reference \cite{BW} uses the Morse potential to produce Robin boundary conditions at an
infinite potential wall, one of the simplest contact interactions, as a limit of a smooth potential.

In this paper we present the mathematical aspects leading to a ``pure'' deformation quantization of the Morse potential (see
eqn. (\ref{MorseV}) below) with general coefficients.

The exponentials in the Morse potential  allow the pseudo-differential form of the $\ast$-eigenvalue equations to be replaced
by a difference-differential equation. (Incidentally, the same is true for any potential that is polynomial in the exponential
function.) The difference-differential equations are then transformed into difference equations that can be solved for all
values of the parameters. The properties of the Mellin and the inverse Mellin transform to relate derivatives with finite
differences are extensively used.  Details are given in the next section, where we also re-derive the known solution for the
special case of a Liouville potential \cite{CFZ}.

In the third section we provide an explicit check to show that the solution provided is indeed consistent with the Wigner
transform of the density matrix of the  Morse wave functions. Our final section is a short conclusion.

\section{Deformation quantization with a Morse\newline potential}

Let us start with a  brief review of Wigner functions and the Wigner-Weyl correspondence (see \cite{ZFC,HWW}, e.g.). For
simplicity, we will restrict to one coordinate $x$ and one conjugate momentum $p$, describing a flat two-dimensional phase space
$\R^2$. The generalization to $\R^{2N}$ is straightforward.

The Wigner function is related to the density operator of canonical quantization. More generally, every operator $\hat Q$ has a
Weyl {\it symbol} $Q(x,p)$ defined by \bea Q(x,p)\ =\ \cW \hat Q\ =\ \frac{1}{(2\pi)^2}\int_{-\infty}^{\infty}d\xi d\eta\ {\rm
Tr}\left[\hat Q\, e^{-i(\xi \hat x+\eta\hat p)}\right] e^{i\xi x+i\eta  p}\qquad \nn\  =\
\frac{1}{2\pi}\int_{-\infty}^{\infty}dy\ e^{-i y p}\ \langle x+\hbar y/2\vert\hat Q\vert x-\hbar y/2\rangle\ .\ \ \ \ \ \ \ \ \
\ \ \ \ \ \ \label{symbol2} \eea The map $\cW:\ \hat Q \mapsto Q(x,p)$, from operators to phase space functions (and
distributions) is called the Wigner transform. It is a homomorphic map from the algebra of operators to the $*$-algebra of
symbols: \ben  \cW(\, \hat Q\,\hat R \,)\ =\ \cW(\hat Q)\,\ast\, \cW(\hat R)\ . \ \ \label{homomW}\een Here the symbols are
multiplied using the Moyal $\ast$-product, \ben \ast\ =\ \exp\left\{\, \frac{i\hbar}2\, \left(\
\stackrel{\leftarrow}{{\partial}_{x}}\stackrel{\rightarrow}{{\partial}_{p}} -
\stackrel{\leftarrow}{{\partial}_{p}}\stackrel{\rightarrow}{{\partial}_{x}}\, \right) \,\right\} \label{Moyal}\een for
consistency with the Wigner transform (\ref{symbol2}). An inverse $\cW^{-1}$ also exists---it is commonly referred to as the
Weyl map: \ben \hat Q\ =:\ {\cW^{-1}}Q(x,p)=\frac{1}{(2\pi)^2}\int_{-\infty}^{\infty}d\xi\ d\eta\ dx\ dp\ Q(x,p) e^{i\xi (\hat
x-x)+i\eta(\hat p-p)}\ . \label{Weylmap} \een Up to normalization, the Wigner function $\rho(x,p)$ is defined as the Wigner
transform of the density operator $\hat\rho$: \ben \rho(x,p)\ :=\ \frac1{2\pi\hbar}\, {\cW} \left(\, \hat\rho\, \right)\
.\label{WigWrho}\een

The Wigner function is a quasi-probability distribution. Expectation values are calculated as \ben  \langle\, Q \,\rangle\ =\
\int dx\,dp\,\, \rho(x,p)\,Q(x,p)\ .\label{expval}\een However, even with a pure state density operator $\hat\rho =
\vert\psi\rangle\langle\psi\vert$, \ben \rho(x,p)\ =\ \frac{1}{(2\pi)^2\hbar}\int_{-\infty}^{\infty}dy\ e^{-i y p}\ \psi(
x+\hbar y/2)\,\psi^*(x-\hbar y/2)\ \ \label{Wigdiag}\een  is negative somewhere (except for the special case of Gaussian wave
functions).

The Wigner function  satisfies the equation of motion
 \ben
i\hbar\,\partial_t \rho(x,p,t)=\left [H,\rho(x,p,t)\right ]_\ast\,,\label{dynam_eqn}
\een
 where $\left[H,\rho\right ]_\ast=H\ast\rho -\rho\ast H$. It can  be expressed as a
 linear combination of stationary Wigner functions with time-dependent coefficients:
\ben
\rho(x,p,t)=\sum_{E_L,E_R}C\sub{E_LE_R}e^{-i(E_L-E_R)t/\hbar}\rho\sub {E_L E_R}(x,p)\ .\label{Wigner _function}
\een
Here $\rho\sub {E_L E_R}= {\cal W}(\vert E_L\rangle\langle E_R\vert )/2\pi\hbar$ denotes the Wigner transform of a matrix element of
the density operator in the energy basis. As $\ast$-eigenfunctions, they can be found by solving the system of equations:
\bea \label{Wigner_function1}
 H\ast\rho\sub {E_L E_R}(x,p)=E_L\ \rho\sub {E_L E_R}(x,p)\ ,\nn \rho\sub {E_L E_R}(x,p)\ast
H=E_R\ \rho\sub {E_L E_R}(x,p)\ .
\eea 
These are known as the $\ast$-eigenvalue (or sometimes
``stargenvalue'') equations.

Alternatively, the Wigner transform \bea\label{Wigner _function3} \rho\sub {E_L E_R}(x,p)=\int_{-\infty}^{\infty}dy\  e^{i y
p}\ \langle x+\hbar y/2\vert E_L
 \rangle \langle E_R \vert x-\hbar y/2\rangle\  \eea allows them to be determined from the wave functions, if known.
 In the case of smooth potentials, the resulting Wigner functions  are known to agree.\footnote{\, For discontinuous
potentials, however, that is not necessarily the case \cite{DP,KW,W,BW}. In \cite{BW} the results reported here are applied to
study the example of the infinite wall, or equivalently, a particle confined to the half-line.}

The goal here is to perform a ``pure'' deformation quantization by solving the $\ast$-eigenvalue equations directly, without
reference to operators or wave functions. This will be done for the Morse potential  \bea V(x)\  =\
\frac{\hbar^2\kappa^2}{2m}\,\left(\, e^{-2\alpha x} - \beta\ e^{-\alpha x} \,\right)\  .\label{MorseV} \eea With its short range
repulsion and longer range attraction, this smooth potential has been useful in many physical applications.   In particular the
Morse potential can be used to recover infinite wall with Robin boundary conditions in deformation quantization \cite{BW}. We
will use the Mellin transform to convert the $\ast$-eigenvalue equations to difference equations.

We use a new method that produces difference equations for potentials that are polynomials of an exponential in
$x$.\footnote{\, Strictly speaking, a difference equation is obtained for any potential that is a linear combination of
exponentials, $\exp(-\alpha_i x)$, $i=1,\ldots,n$, say. That is not likely to be helpful, however, unless all the ratios
$\alpha_i/\alpha_j$ are rational.} This is significant since solutions to the $\ast$-eigenvalue equations are generally
difficult to find (see \cite{CFZ}, e.g.). To see why, let us start with a more general Hamiltonian and then specialize to the
Morse potential. Writing \ben E_L\ =:\ \frac{ \hbar^2 k_L^2}{2m}\ ,\ \ \ E_R\ =:\ \frac{ \hbar^2 k_R^2}{2m}\ ;
\label{ELkLERkR}\een the $\ast$-eigenvalue equations (\ref{Wigner_function1}) are of infinite order in
momentum-derivatives for a generic Hamiltonian $H=p^2/2m+V(x)$:
\ben
\left(\frac{p^2-\hbar^2 k^2_L}{2m}\right)\rho-\frac{\hbar
^2}{8m}\partial^2_x\rho-\frac{i\hbar p}{2m}\partial_x\rho+\sum^{\infty}_{n=0}\frac{1}{n!} \left(\frac{i\hbar
}{2}\right)^n\partial^n_x V\partial^n_p\rho=0\ ,\ \ \ \ \ \ \label{diff_left}
\een
and similarly for the ``right'' equation.
 However, the exponential form of the Morse potential (\ref{MorseV}) allows them to be written as differential-difference
 equations--the
 exponentials generate translations in the momentum. Their explicit form becomes
\bea\label{stargen_exp_l} \frac{\hbar ^2}{8m}\partial^2_x\rho+\frac{i\hbar p}{2m}\ \partial_x\rho=\,\ \ \ \ \ \ \ \ \ \ \ \ \ \ \ \ \
\ \ \ \ \ \ \ \ \ \ \ \ \ \ \ \ \ \ \ \ \ \ \ \ \ \ \ \ \ \ \ \ \ \ \ \ \ \ \ \ \ \ \ \ \ \ \\  \nonumber \frac{\hbar^2\kappa^2}{2m} e^{-2\alpha
x}\rho(x,p-i\hbar \alpha)-\frac{\beta\hbar^2\kappa^2}{2m} \ e^{-\alpha x}\rho\left(x,p-\frac{i\hbar
\alpha}{2}\right)+\left(\frac{p^2-\hbar^2 k^2_L}{2m}\right)\rho\ \ \nonumber \eea and  \bea\label{stargen_exp_r} \frac{\hbar
^2}{8m}\partial^2_x\rho-\frac{i\hbar p}{2m}\partial_x\rho=\ \ \ \ \ \ \ \ \ \ \ \ \ \ \ \ \ \ \ \ \ \ \ \ \ \ \ \ \ \ \ \ \ \ \ \ \ \ \
\ \ \ \ \ \ \ \ \ \ \ \ \ \ \ \ \ \ \ \ \ \ \ \ \ \ \\ \nonumber \frac{\hbar^2\kappa^2}{2m} e^{-2\alpha x}
\rho(x,p+i\hbar \alpha)-\frac{\beta\hbar^2\kappa^2}{2m}\
e^{-\alpha x}\rho\left(x,p+\frac{i\hbar \alpha}{2}\right)+\left(\frac{p^2-\hbar^2 k^2_R}{2m}\right)\rho\ . \nonumber \eea

The  integral transform technique leads to further simplifications. Suppose the Wigner function can be written as
 \ben
 \rho(x,p)\ =\ R(u,p)\ ,\ \ u\,:=16 e^{4\alpha x}\al^4/\kappa^4\,. \label{rhoR}
\een  The Mellin transform of the Wigner function is
 \ben
 W(s,p)\ :=\ {\cal M} \{R \}(s,p)\
=\ \int_{0}^{\infinity}\, u^{s-1}\, R(u,p)\, du\ . \ \label{Wxpdef} \een To transform (\ref{stargen_exp_l})  and
(\ref{stargen_exp_r}) into difference equations for $W(s,p)$ we consider the inverse Mellin transform
\ben
R(u,p)\ =\ \frac 1
{2\pi i}\, \int_{c-i\infinity}^{c+i\infinity}\, u^{-s}\, W(s,p)\, ds\ , \label{RzpWinv}
\een
 where the constant $c$ can be any
constant for which the transform converges according to the Mellin inversion theorem.

\subsection{Solution for the Liouville potential}

We will first apply the method to the Liouville potential \ben
 V_L(x)\ =\ \frac{\hbar^2\kappa^2}{2m}\, e^{-2\al x}\ . \label{VLiou}
 \een
 It is the simplest case as it can be viewed as a Morse potential for $b=0$. Also, we can check our results
 since the Wigner functions for this
 potential have already  been found in \cite{CFZ}.

The $\ast$-eigenvalue equations (\ref{Wigner_function1}) imply the difference equations \bea (p/\hbar
+2i\alpha s)^2 W_0(s,p) + (2\alpha)^2 W_0(s- 1/2,p-i \alpha\hbar) \ =\ k_L^2 W_0(s,p)\ ,\nn
 (p/\hbar -2i\alpha s)^2 W_0(s,p) + (2\alpha)^2 W_0(s-1/2,p+i \alpha\hbar) \ =\ k_R^2 W_0(s,p)\ . \label{starWL}
\eea Let us now assume that the solution is factorized into two parts -- ``left'' and ``right'' factors -- each depending on
$k_L$ or $k_R$ only: \ben W(s,p)\ =\ N\, w_{L}\big(s- \frac{ip}{2 \alpha\hbar},\, k_L\big)\, w_{R}\big(s+ \frac {ip}{2
\alpha\hbar},\, k_R \big)\,\ ,\label{WwLwR}
\een
 with $N$ a normalization constant.  For the left factor we find
\bea\label{starwMsiL}
(p/\hbar+2i \alpha s)^2 \,w_L(s-ip/2 \alpha\hbar) + (2\alpha)^2 \,w_L(s- ip/2\al \hbar-1)=\ \ \ \ \ \ \ \ \ \ \ \\
\nonumber \ \ \ \ \ \ \ \ \ \ \ \ \ \ \ \ \ \ \ \ \ \ \ \ \ \ \ \ \ \ \ \ \ \ \ \ \ \ \ \ \ \ \ \ \ \  =k_L^2\, w_L(s - ip/2
\al\hbar)\, \nonumber
\eea
 and for the right factor
\bea
\label{starwMsiR} (p/\hbar-2i \alpha s)^2 \,w_R(s+ip/2 \alpha\hbar) +
(2\alpha)^2 \,w_R(s+ ip/ 2\al\hbar -1)=\ \ \ \ \ \ \ \ \ \ \ \\ \nonumber \ \ \ \ \ \ \ \ \ \ \ \ \ \ \ \ \ \ \ \ \ \ \ \
\ \ \ \ \ \ \ \ \ \ \ \ \ \ \ \ \ \ \ \ \ \  =k_R^2\, w_R(s + ip/2 \al\hbar)\ .\nonumber \eea
 Using the substitution $t=s-ip/2\al\hbar$ we arrive at
\ben w(t-1,k_L)\ =\ \left[ t^2 + \frac{k_L^2}{(2\alpha)^2} \right]\,w(t,k_L)\ . \label{wzdiff} \een Equation (\ref{starwMsiR})
also leads to the above equation if we use $t=s+ip/2\al\hbar$ and $k_R$ instead.  Therefore we need only work with
(\ref{wzdiff})  and the solutions will just differ in their arguments and  labels of $k$.

The solution of (\ref{wzdiff})  is \ben w(t,k_L)=\Gamma(-t+ik_L/2\al )\,\Gamma(-t-ik_L/2\al )\ , \label{Liouville_diff}
\een by the defining property $\Gamma(z+1)=z\Gamma(z)$ of the gamma function. Tracing back to equations
(\ref{WwLwR}-\ref{starwMsiL}), (\ref{RzpWinv}) and (\ref{rhoR}), we can write the Wigner function in terms of the inverse
Mellin transform:
\bea
\rho\sub {k_L k_R}(x,p)\propto\int_{c-i\infinity}^{c+i\infinity}ds\ u^{-s} \times\ \ \ \ \ \ \ \ \ \
\ \ \ \ \ \ \ \ \ \ \ \ \ \ \ \ \ \ \ \ \ \ \ \ \ \ \ \ \ \ \ \  \\ \nonumber \
\ \ \ \prod_{\pm,\pm'}\Gamma\left(-s+\frac{i(p/\hbar\pm k_L)}{2\al }\right)\Gamma\left(-s-\frac{i(p/\hbar\pm'
k_R)}{2\al }\right).\label{int_rep_Wigner_Liouv}
\eea
 This last is an integral representation of the Meijer $G$-function.
Using equation (43) on pg. 353 in \cite{Tables}  the Wigner function becomes \bea \rho\sub {k_L k_R}(x,p)\propto \ \ \ \
\ \ \ \ \ \ \ \ \ \ \ \ \ \ \ \ \ \ \ \ \ \ \ \ \ \ \ \ \ \ \ \ \ \ \ \ \ \ \ \ \ \ \ \ \ \ \ \ \ \ \ \ \ \ \ \ \ \ \ \ \ \ \ \
\ \ \ \ \ \ \ \ \ \ \ \\ \nonumber G^{40}_{04}\left( \frac{1}{u}\biggl\vert\frac{i(p/\hbar+k_L)}{2\al
},\frac{i(p/\hbar-k_L)}{2\al },-\frac{i(p/\hbar-k_L)}{2\al }, -\frac{i(p/\hbar+k_L)}{2\al }\right)\,,\label{Meijer_Liouville}
\eea where we used the identity \ben G^{40}_{04}\left( u\vert 1-a_1, 1- a_2, 1-a_3, 1-a_4\right)=G^{40}_{04}\left( 1/u\vert a_1,
a_2, a_3, a_4\right)\,.\label{Meijer_identity} \een As it should, the formula for the Wigner function (\ref {Meijer_Liouville})
coincides with the one obtained using different methods  in \cite{CFZ}. It describes the phase-space quasi-distribution for a
particle in a Liouville potential. The advantage of the method proposed here is that it can be generalized to the Morse
potential.

\subsection{Solution for the Morse potential}
Now let us go back to the original problem of finding the Wigner function for the potential (\ref{MorseV}).  The left $\ast$-eigenvalue  equation  has the
form \bea\label{starWM} (p/\hbar+2i\alpha s)^2
W_b(s,p) + (2\alpha )^2 W_b(s- 1/2,p-i\alpha\hbar)  \ \ \ \ \ \ \ \ \ \ \ \ \ \ \ \ \ \ \ \ \ \ \ \ \ \ \ \ \\
\nonumber
 -\frac b 2 (2\al )^2 \ W_b(s- 1/4,p- i\alpha\hbar/2)\ =\  k_L^2\, W_b(s,p)\ , \nonumber
\eea where we set $b=\beta\kappa/\al $ . We also have  the complex conjugate (right)  equation, with $k_L$ replaced by $k_R$. To solve the new left difference
equation (\ref{starWM}), we substitute the ansatz (\ref{WwLwR})  to obtain \bea\label{starwMsi} (p/\hbar+2i\alpha s)^2
\,w_L(s-ip/2\alpha\hbar, k_L) + (2\alpha )^2 \,w_L(s- ip/2\al\hbar -1, k_L)  \ \ \ \ \ \ \ \ \ \ \ \
\\ \nonumber - \frac b 2 (2\alpha )^2 \,w_L(s-ip/2\al\hbar - 1/2, k_L)\ =\ k_L^2\, w_L(s - ip/2\al\hbar, k_L)\ .
\nonumber
 \eea
  Using the same substitution  $t=s-ip/2\al\hbar$ as in (\ref{starwMsiR}) and switching to $w_b=w_L$ to
account for  the parameter dependence, we arrive at the difference equation relevant to the Morse potential: \ben
 w_b(t-1,
k_L)-\frac{b}{2}\, w_b(t-1/2, k_L)\ =\ \left[ t^2 + \frac{k^2_L}{(2\alpha )^2} \right]\, w_b(t,k_L)\ .
\label{wsidiff}
 \een
  This equation has a trivial solution for $b=0$,  the Liouville case.  The right factor satisfies
  an identical equation with $k_L$ replaced by $k_R$ and $ t=s+ip/2\al\hbar$.

For $b=1$ the solution can be written in terms of gamma functions, as in  the $b=0$ case
\bea\label{Diff_CFZ_case}
 w_1(t,k)\propto\Gamma\left(-t+\frac{ik}{2\al }\right)\Gamma\left(-t+\frac{1}2-\frac{ik}{2\al }\right)+\ \ \
 \ \ \ \ \ \ \ \ \ \ \ \ \ \ \ \ \ \ \ \ \ \ \ \ \ \ \ \ \ \ \\ \nonumber
 \Gamma\left(-t+\frac{1}2 +\frac{ik}{2\al }\right)\Gamma\left(-t-\frac{ik}{2\al }\right) . \nonumber
\eea The inverse Mellin transform then gives  us the Wigner function:
\bea\label{w_1}
 \rho\sub {k_L k_R}(x,p)\propto\ \ \ \ \ \
\ \ \ \ \ \ \ \ \ \ \ \ \ \ \ \ \ \ \ \ \ \ \ \ \ \ \ \ \ \ \ \ \ \ \ \ \ \ \ \ \ \ \ \ \ \ \ \ \ \ \ \ \ \ \ \ \ \ \ \ \ \ \ \
\ \ \ \ \ \ \  \\ \nonumber G_{04}^{40}\left(\frac{1}{u}\biggr\vert\frac{ip}{2\al\hbar}+\frac{ik_L}{2\al },\frac 1 2
+\frac{ip}{2\al\hbar}-\frac{ik_L}{2\al },-\frac{ip}{2\al\hbar}+\frac{ik_R}{2\al },\frac 1
2-\frac{ip}{2\al\hbar}-\frac{ik_R}{2\al }\right)+\\ \nonumber
G_{04}^{40}\left(\frac{1}{u}\biggr\vert\frac{ip}{2\al\hbar}+\frac{ik_L}{2\al },\frac 1 2+\frac{ip}{2\al\hbar}+\frac{ik_L}{2\al
},\frac 1 2-\frac{ip}{2\al\hbar}+\frac{ik_R}{2\al },-\frac{ip}{2\al\hbar}-\frac{ik_R}{2\al }\right)+\\  \nonumber
G_{04}^{40}\left(\frac{1}{u}\biggr\vert\frac 1 2+\frac{ip}{2\al\hbar}+\frac{ik_L}{2\al
},\frac{ip}{2\al\hbar}-\frac{ik_L}{2\al},-\frac{ip}{2\al\hbar}+\frac{ik_R}{2\al },\frac 1
2-\frac{ip}{2\al\hbar}-\frac{ik_R}{2\al }\right)+\\ \nonumber G_{04}^{40}\left(\frac{1}{u}\biggr\vert\frac 1
2+\frac{ip}{2\al\hbar}+\frac{ik_L}{2\al },\frac{ip}{2\al\hbar}-\frac{ik_L}{2\al },\frac 1
2-\frac{ip}{2\al\hbar}+\frac{ik_R}{2\al },-\frac{ip}{2\al\hbar}-\frac{ik_R}{2\al }\right).\ \ \nonumber \eea

Another solution that is easy to find is for $b=2$:\,\footnote{\, The $b=2$ Wigner function can also be found from the
Liouville case using supersymmetric quantum mechanics. In deformation quantization, the ladder operators of supersymmetric
quantum mechanics are replaced by functions and a star product is used, however the transition is fairly straightforward
\cite{CFZ}.} \bea
 w_2(t,k)\propto(t+1/4)\prod_{\pm}\Gamma\left(-t\pm\frac{ik}{2\al }\right)-\prod_{\pm}\Gamma\left(-t+\frac{1}2
 \pm\frac{ik}{2\al }\right). \label{vdef}
\eea
 It can be written in a different form which shows a pattern shared with the case $b=1$:
\bea\label{ell_1_new_form}
 w_2(t,k)\propto(2ik/\al +1)\,\Gamma\left(-t+ik/2\al \right)\Gamma\left(-t+1-ik/2\al \right)+\\ \nonumber
\frac{4ik}{\al} \ \Gamma\left(-t+1/2+ik/2\al \right)\Gamma\left(-t+1/2-ik/2\al \right) +\nonumber \\
 (2ik/\al -1)\,\Gamma\left(-t+1+ik/2\al \right)\Gamma\left(-t-ik/2\al \right).\ \ \nonumber
\eea The Wigner function can be found with a trivial but lengthy calculation that is essentially identical to the $b=0$ and
$b=1$ cases. The exact combination of Meijer $G$-functions is not of interest to us; we will derive a general expression that
includes this one later.

A useful observation is that the left factors  (for different $b$) can be written as: \ben w_1=C_1 w_0(t-1/4,k+i\al
/2)+C_2w_0(t-1/4,k-i\al /2)\label{w1rewrite} \een and \ben w_2=C_1 w_0(t-1/2,k+i\al )+C_2 w_0(t-1/2,k)+C_3 w_0(t-1/2,k-i\al )\
.\label{w2rewrite}
 \een
This suggests that by choosing the constants correctly we can write the solution for any integer $b$ as
 \bea
 w_b(t,k)\propto\sum_{n=-\frac{b}{2},-\frac{b}{2}+1,...,\frac{b}{2}} C_n^b\,w_0\left(t-b/4,k-i n\al \right).\label{ansatz}
\eea We can substitute this ansatz into the equation  (\ref{wsidiff})  using undetermined coefficients.  In principle,
comparison of  the coefficients   of  independent terms can  determine $ C_n^b$ for any $b$. This seems to fail, however, in
the case of non-integer $b$. Furthermore, even for the simplest cases this program is  very difficult to carry out.\footnote{\,
For integer $b$, see eqn. (\ref{coeffs}) below, however.} Clearly we need  an algorithm that reproduces the constants directly
and allows a generalization to include all Morse potentials of the form (\ref{MorseV}).

\subsection{Systematic solution of the difference equations}

We now show how to find the relevant solutions of the $\ast$-eigenvalue equations for the Morse potential, for all $b\in\R_+$.
We exploit once again the property of the Mellin transform to relate differential equations and their solutions to difference
equations and their solutions.

To convert our difference equation (\ref{wsidiff})  into a differential equation we use the following two properties of the
Mellin transform:
\bea\label{Mellin_properties1}
 {\cal M}\{\tau^2 f''(\tau)+\tau f(\tau)\}(s)=s^2{\cal M}\{f(\tau)\}(s)\,, \\
 {\cal M}\{\tau^a f(\tau)\}(s)={\cal M}\{f(\tau)\}(s+a)\,.\label{Mellin_properties2}
\eea
 If we apply the inverse Mellin transform directly to  (\ref{wsidiff}), we end up with an equation that we cannot solve.
This is because the Mellin transform converts argument translations into powers of the argument via (\ref{Mellin_properties2}).
To eliminate fractional powers,  we  use the substitution $s=2t$. The new equation for $\tilde w_b(s)=w(t(s))$
 \bea
  \tilde w_b(s-2)-\frac{b}{2}\tilde w_b(s-1)=\left[\left(\frac s 2\right)^2+\frac{k^2}{(2\al )^2}\right]\tilde
w_b(s) \label{difference_new}
\eea
 results in a simpler, integrable equation:
 \bea
 \tau^2 f''(\tau)+\tau
f'(\tau)+\bigl[\left(k/\al \right)^2-4/\tau^2+2b/\tau\bigr] f(\tau)\,,\label{underlying}
\eea
 where $\tilde w(s)={\cal M}\{f(\tau)\}(s)$.

 The solution $f(\tau)$ of this equation can be found if we make  the substitution
$f(\tau)=\tau^{-1/2}g(\tau)$ and then $u(z)=g(t(z))$, where $z=1/\tau$. The new function $u(z)$ satisfies the so-called
Whittaker equation, treated in \cite{WW}, Chapter XVI, and  also in the Appendix.  Its two linearly independent solutions are
defined in \cite{Handbook}, pg. 755.
 They are called Whittaker functions and can be
expressed in terms of the Tricomi confluent hypergeometric function $U(\mu,\nu,z)$ and the Kummer confluent hypergeometric
function $M(\mu,\nu,z)$:\,\footnote{\, The Whittaker function $M_{lm}(z)$ should not be confused with the Kummer function
$M(\mu,\nu,z)$ in the above equation. Subscripts are used to denote the parameters of the Whittaker functions in the
literature, and the explicit bracket notation is used for confluent hypergeometric functions. For further information involving
the hypergeometric functions see \cite{Handbook}, p.753 and \cite{Abram_Stegun}, p.503-506. } \bea
&M_{lm}(z)=z^{m+1/2}e^{-z/2}\,M( 1/2+m-l,1+2m;z)\ ,  \label{Whitt_M} \\
&W_{lm}(z)=z^{m+1/2}e^{-z/2}\,U(1/2 +m-l,1+2m;z)\ .\label{Whitt_W} \eea For our purposes, we only need the definitions of those
functions \bea
M(\mu,\nu;z)=\sum_{n=0}^\infty \frac{(\mu)_n}{(\nu)_n}\frac{y^n}{n!}\ ,\qquad\quad \label{Kummer_M} \\
U(\mu,\nu;z)=\frac{\Gamma(\nu-1)}{\Gamma(\mu)} z^{1-\nu} M(1+\mu-\nu,2-\nu;z)\quad \nn +\
\frac{\Gamma(1-\nu)}{\Gamma(\mu-\nu+1)}M(\mu,\nu;z)\ .\label{Tricomi} \eea Here we use the Pochhammer symbol
$(\mu)_n:=\mu(\mu+1)...(\mu+n-1),\ (\mu)_0:=1$.

The solution is: \bea f(\tau)=\tilde C_1\tau^{1/2}M_{\frac b 2,\frac{ik}{\al}}(4/\tau)+ \tilde C_2\tau^{1/2}W_{\frac b
2,\frac{ik}{\al}}(4/\tau)\ .\label{solution_beta} \eea This is the general solution and it therefore depends on two arbitrary
constants, $\tilde C_1,\, \tilde C_2$. We must set $\tilde C_1=0$ to describe the physical states, however. To see that we have
to transform back to the solutions of the difference equations and compare with the known solutions, re-derived in the previous
sections.

Let us first confirm that (\ref{solution_beta}), with $\tilde C_1=0$, indeed recovers the known solutions
(\ref{Liouville_diff}, \ref{Diff_CFZ_case}, \ref{ell_1_new_form}) for $b=0,1,2$,  respectively, and that it also justifies the
ansatz (\ref{ansatz}) for all non-negative integer $b$. For $b\in\N_0$ we can write the Whittaker $W$-function in terms of the
modified Bessel functions \bea\label{Whittaker_Bessel}
W_{\frac{n}{2},\,\mu}(y)=\frac{y^{\frac{n+1}{2}}}{\sqrt \pi}\left(\frac{1-n}{2}+\mu\right)_n\times\  \QQ\qq\qq\\
\label{repn_of_Whitt}
\qq\qq\times\sum_{k=0}^{n}\frac{(-1)^{n+k}(2k-n+2\mu)(-n)_{n-k}}{\Gamma(n-k)(k-n+2\mu)_{n+1}}K_{-k+\frac{n}{2}-\mu}\left(\frac{y}2\right)\
. \nonumber \eea With the help of the integral representation of the Bessel functions (effectively finding the inverse Mellin
transform), \bea K_\nu(z)=\frac{1}{4\pi i}\int^{c+i \infty}_{c-i \infty}\Gamma(s)\Gamma(s-\nu)\left(\frac z 2\right)^{\nu-2s}
ds\ ,\label{integral_representation_K} \eea we can find the solution of (\ref{difference_new}) for integer $b$:
 \bea
 w_b(t,k)\propto\sum_{n=0}^b C_n^b\,\Gamma\left(-t+n/2+ik/2\al \right)\Gamma
 \left(-t-n/2+b/2-ik/2\al \right).\label{full_solution}
\eea This is nothing more than (\ref{ansatz}) with a shifted summation index.  The coefficients are now explicit, however: \ben
C_n^b=\frac{(-1)^{n}(2n-b+2ik/\al)(-b)_{b-n}}{(b-n)!(n-b+2ik/\al)_{b+1}}\ .\label{coeffs} \een

The Wigner function for $b\in\N_0$ can be found from (\ref{full_solution}) as in the preceding two subsections, with result
\bea\label{G} \rho\sub {k_L k_R}(x,p)=\sum_{m,n=0}^{b} C_m^b  C_n^b \times \ \ \ \ \  \ \ \  \  \ \ \ \ \ \ \ \ \ \ \ \ \ \ \ \
\ \ \ \ \ \ \ \ \ \ \ \ \ \ \ \ \ \ \ \ \  \ \ \ \ \ \ \ \ \ \ \ \  \\  \nonumber \times
G^{40}_{04}\left(\frac{1}{u}\right\vert\frac{n}{2}+\frac{i(p/\hbar+k_L)}{2\al},\frac{b}{2}-\frac{n}{2}+\frac{i(p/\hbar-k_L)}{2\al
},\ \ \ \ \ \ \ \ \ \ \ \ \ \ \ \ \ \ \ \ \ \   \ \ \ \ \ \  \\ \nonumber \frac{m}{2}-\frac{i(p/\hbar-k_R)}{2\al
},\frac{b}{2}-\frac{m}{2}-\frac{i(p/\hbar+k_R)}{2\al }\biggr)\ .\nonumber \eea

Bound states can also be treated this way; one only needs to consider imaginary $k$, for energies $E<0$. The  resulting form of
the Wigner function  differs, however, from the expression found by the Wigner transform
 \ben
 \rho(z,p)\propto z^{2\nu-b+1}\sum^{\nu}_{l_1,l_2=0}{{b-\nu-1} \choose {\nu-l_1}} {{b-\nu-1} \choose {\nu-l_2}}\frac{(-z)^{l_1+l_2}}{{l_1!\ l_2!}}
 \ K_{l_1-l_2-2ip/\al}(z)\ .\label{Wigner_bound_final}
 \een
 To write it in this form we use the properties of the Whittaker functions. Recall that the energies are given by
 (\ref{energy_Morse}) and we can write \ben
f(\tau)\propto\tau^{1/2}W_{{\frac b 2},{\frac{ik}{\al}}}(4/\tau)=\tau^{1/2}W_{{\frac b 2},{\frac b 2 -\nu -\frac 1 2}}(4/\tau).
\een The relationship \ben W_{{a},{a -\nu -\frac 1 2}}(z)\ =\ (-1)^\nu\nu !\,
z^{a-\nu}e^{-z/2}L^{2a-2\nu-1}_{\nu}\label{Whittaker_special_case} \een (for integer $\nu$) and (\ref{Laguerre_assoc}) allow us
to write the solution as \ben f(\tau)\propto e^{-2/\tau}\sum^{\nu}_{l=0}\frac{(-1)^l}{l!} {{b-\nu-1}\choose{\nu-l}}\left(\frac
4 \tau\right)^{l+b/2-\nu-1/2}.\label{bound} \een Transforming the above we find the corresponding factor: \ben w(t)\propto
\sum^{\nu}_{l=0}\frac{ (-2)^l 2^{2t}}{l!} {{b-\nu-1}\choose{\nu-l}} \Gamma\left(-2t+l+b/2-\nu-1/2 \right).\label{bound_factor}
\een With the help of the inverse Mellin transform we find (\ref{Wigner_bound_final}) as in the unbound case, using (\ref{integral_representation_K}) and (\ref{RzpWinv}).

Let us  proceed to the case of non-integer $b$. For $b\in\R$, we find an
explicit and closed expression for the solution of the difference equation. We start by rewriting the solution of the
differential equation (\ref{underlying}) in terms of hypergeometric functions: \bea f(\tau)=e^{-2/\tau}\tau^{-ik/\al
}U\left(\frac 1 2 -\frac b 2 +\frac{ik}{\al},1+\frac{2ik}{\al };\frac 4 \tau\right)\ .\label{solution_Tricomi_Kummer} \eea
 This is necessary  in order to perform the inverse Mellin transform of $f(\tau)$ in closed terms, which
 presents a technical problem if we use the Whittaker function.

We use the relationship (\ref{Tricomi})  between the Kummer and Tricomi hypergeometric function and the integral expression
\bea e^{-\sigma x}M(\beta,\gamma;\lambda x)=\int_{c-i\infty}^{c+i\infty}ds\,\, x^{-s}\sigma^{-s}\,\Gamma(s)\,{\,
}_{2}F_{1}(\beta,s\,;\gamma\,;\lambda \sigma^{-1})\label{Kummer_Gauss_transform} \eea to  make it possible to find the inverse
Mellin transform of (\ref{solution_Tricomi_Kummer}). It is given in  term of the Gauss hypergeometric function $_{2}F_{1}$.
Switching back to our original variable $t$, and indicating the $k$-dependence explicitly, we can write \bea w_b(t,k)\propto
\frac{4^{t+ik/2\al }\Gamma(-2ik/\al )}{\Gamma(1/2-b/2-ik/\al )}\Gamma(-2t+ik/\al )\times\ \ \ \ \ \ \ \ \ \ \ \ \ \ \ \ \ \ \ \
\ \ \ \ \ \nn _{2}F_{1}\left(1/2-b/2+ik/\al ,-2t+ik/\al ;1+2\,ik/\al ;2\right)+\nn \frac{4^{t-ik/2\al }\Gamma(2ik/\al
)}{\Gamma(1/2-b/2+ik/\al)}\Gamma(-2t-ik/\al )\times\ \ \ \ \ \ \ \ \ \ \ \ \ \ \ \ \ \ \ \ \ \ \ \ \ \nn
_{2}F_{1}\left(1/2-b/2-ik/\al ,-2t-ik/\al;1-2\,ik/\al ;2\right)\,.\,\label{full_solution2} \eea This is the solution of the
difference equation  (\ref{wsidiff}) for any real $b$, including the ones we already found. After substitution in
(\ref{rhoR},\,\ref{RzpWinv},\,\ref{WwLwR}), it yields our main result: \bea\label{Wigner_function_complete} \rho\sub {E_L
E_R}(x,p)\ \propto\QQ\QQ\QQ\nn \qq\int_{c-i\infinity}^{c+i\infinity}ds\ u^{-s}
w_b\left(s-\frac{ip}{2\al\hbar},k_L\right)\, \, w_b\left(s+\frac{ip}{2\al\hbar},k_R\right)\,\ . \eea

\section{Wigner functions from wave functions for a Morse  potential}

For completeness, let us confirm that our solutions of the $\ast$-eigenvalue equations (\ref{Wigner_function1}) are the same as
the Wigner functions derived  from the  wave functions (see the Appendix) using (\ref{Wigner _function3}).

Consider the unbound states first
\bea
\psi(y)\ =\ Ce^{-y/2}\,\tilde A \, y^{i k/\alpha}\, M\left( \frac1 2-\frac{b }{2 } + \frac{i k}{\alpha}, 1+ \frac{2 i k}
{\alpha} ; y \right)+\nonumber\\
+\ Ce^{-y/2}\, \tilde A^*\, y^{-i k/\alpha}\, M \left(\frac1 2-\frac{b }{2 } -\frac{i k}{\alpha}, 1- \frac{2 i
k}{\alpha}; y\right)\,. \ \label{MatsumotoWF} \eea Using  the substitutions  $w=e^{-\al\hbar  y/2}$ and $v=2\kappa e^{-\al x}/\al$ we
can rewrite the integral transform (\ref{Wigner _function3}).  The result involves integration over $w$ of products of the type \ben
 \exp{[-v(w+w^{-1})/2] }w^m v^n M\left(a_1, b_1 ; v\,w \right) M\left(a_2, b_2; v/w \right)\,,\nonumber
\een which (to the best of our knowledge) are not integrable in closed form. However we can expand the Kummer $M$-function as
in (\ref{Kummer_M}). Then all the integrations can be performed explicitly  using the integral representation of the Bessel
$K$-function \ben K_\nu(z)=\frac 1 2\int_0^\infty dw\ w^{-(\nu+1)}e^{-\frac{1}{2}z\left(w+1/w\right)} \,. \label{K_integral}
\een This procedure leads to an infinite sum: \bea \label{Wigner_final_any}
 \rho(v,p)\propto\sum_{m,n=0}^{\infty}\frac{v^{m+n}}{m!n!}\biggl [\ \tilde A^2 v^{2ik /\al}\frac{(\chi)_m
(\chi)_n}{(\varsigma)_m (\varsigma)_n}K_{n-m-2ip/\hbar\alpha}(v)+ \\ \nonumber \vert \tilde A\vert^2\frac{(\chi)_m (\bar
\chi)_n}{(\varsigma)_m (\bar \varsigma)_n}K_{m-n+2i(k-p/\hbar)/\alpha}(v)+\\ \nonumber
 \vert \tilde A\vert^2\frac{(\bar \chi)_m ( \chi)_n}{(\bar \varsigma)_m ( \varsigma)_n}K_{m-n-2i(k+p/\hbar)/\alpha}(v)+ \\ \nonumber
(\tilde A^*)^2 v^{-2ik /\al }\frac{(\bar \chi)_m (\bar \chi)_n}{(\bar \varsigma)_m (\bar
\varsigma)_n}K_{n-m-2ip/\hbar\alpha}(v)\biggr ]\,,\nonumber
 \eea
with $\chi:=1/2-b/2+ik/\al $ and $\varsigma := 1+2 i k/\al $. This expression can now be used to compare with
(\ref{full_solution2},  \ref{Wigner_function_complete}).

To do that we need to calculate explicitly the contour integral (\ref{Wigner_function_complete}). We are facing a similar
problem --- the integrand is too complicated and we need to rewrite  the Gauss hypergeometric function as \ben
\label{def2F1} {}_2F_1(a,b;c;z)\ =\ \sum_{n=0}^\infty \frac{(a)_n\,(b)_n}{(c)_n}\, \frac{z^n}{n!}\ \een in order to integrate
in closed form. With the use of the contour representation of the  Bessel $K$-function (\ref{integral_representation_K}) and
some simple algebra we recover the infinite sum (\ref{Wigner_final_any}). This confirms that the Wigner function
(\ref{Wigner_function_complete}) indeed coincides with the Wigner transform of the density matrix for the Morse potential
from the Schr\"odinger treatment.

The Wigner functions of the  bound states can be obtained from the known wave functions (\ref{Morse_bound_pre}) using the
integral transform  (\ref{Wigner _function3}).\footnote{\, For $b=2$, the  bound-state Wigner functions for the Morse potential
(\ref{MorseV}) have already been treated this way in \cite{FRW}. However, they are not obtained there by solving their
dynamical equations, as we have done. In addition, while we find the unbound-state Wigner functions in the same way, the
unbound states are not considered in \cite{FRW}.} Equation (\ref{Laguerre_assoc}) allows us to evaluate the integral in closed
form using the modified Bessel functions $K_\nu(x)$. The substitutions $v=2\kappa e^{-\al x}/\al$  and $w =\exp(-\al \hbar y/2)$ in
(\ref{Wigner _function3}) and the integral representation (\ref{K_integral}) are used.

\section{Conclusion}

Our main result is the solution (\ref{Wigner_function_complete}, \ref{full_solution2}) of the $\ast$-eigenvalue equations
(\ref{Wigner_function1}) for the Morse potential (\ref{MorseV}) with arbitrary real $b$. It subsumes the simpler formula
(\ref{Wigner_function_complete},\,\ref{ansatz},\,\ref{coeffs}) valid for all $b\in\N_0$.

The solutions obtained have already been applied to a study of Robin boundary conditions in phase-space quantum mechanics
\cite{BW}.

It should also be possible to use our method to solve the $\ast$-eigenvalue equations for other potentials that are polynomial
in an exponential (say $\exp(-\alpha x)$).

\vfill\eject\begin{tabbing}{\bf\large Appendix:}\qquad \ \= {\bf\large The Morse potential in Schr\"odinger}\\ \> {\bf\large
quantum mechanics}\end{tabbing}

Following Matsumoto \cite{M}, we can solve the stationary Schr\"odinger equation for the unbound wave functions of the Morse
potential (\ref{MorseV}). The substitution $\psi(x) = \phi(z)$, $z=\exp(-\alpha x)$, changes the Schr\"odinger equation into
\ben z^2 \phi'' +z \phi'+\frac{1}{\alpha^2}\left[ \frac{2mE}{\hbar^2}-\kappa^2 z^2 +\kappa^2 \beta\ z\right]\phi=0\
.\label{Morse_schrod} \een This can be further transformed into canonical form (without a first derivative term) using the
substitution $\phi(z)\ =\ z^{-1/2}\,F(z)$. Changing the variables to $y\ :=2\kappa z/ \al$ leads to the Whittaker equation,
treated in \cite{WW}, Chapter XVI: \ben
 f''\ +\ \left\{ -\frac 1 4\ +\ \frac{b}{2 }\frac 1 y\ +\ \frac 1{y^2}
 \left[ \frac 1 4 -\left(\frac{i k}{\alpha}\right)^2\right] \right\}\, f\ =\ 0\ , \label{Whittaker}
\een where $f(y):=F(\alpha y/2\kappa)$ and, as before, $k =\sqrt{2 m E}/\hbar$ and $b=\beta \kappa/\al$.

Now the wave function can be written as \ben \psi_k(x)=e^{\al x/2}\left[\tilde C_1 M_{\frac{b }{2 },\frac{i
k}{\alpha}}(y(x)) + \tilde C_2 W_{\frac{b}{2} ,\frac{i k}{\alpha}}(y(x))\right]\,.\label{wave_Fn} \een Imposing
reality yields $\tilde C_1=0$. The second term has physical asymptotic behaviour: for large positive $x$ it is sinusoidal with
a phase depending on the potential parameters; for negative $x$ far from the origin, there is the expected rapid exponential
decay of a classically forbidden region. The wave function is therefore \ben \psi_k(x)=C e^{\al x/2}
W_{\frac{b }{2 },\frac{i k}{\alpha}} \left(\frac{2\kappa }{\al}e^{-\al x}\right)\ .\label{wave_Fn_final} \een

With the help of equation (\ref{Tricomi}) we can rewrite this  result in a form similar to that given by Matsumoto in \cite{M}
for a Morse potential with $b=2$. The wave function is manifestly real in this form: \bea \psi(y)\ =\ Ce^{-y/2}\,\tilde A \,
y^{i k/\alpha}\, M\left( \frac1 2-\frac{b }{2 } + \frac{i k}{\alpha},
1+ \frac{2 i k}{\alpha} ; y \right)+\nonumber\\
+\ Ce^{-y/2}\, \tilde A^*\, y^{-i k/\alpha}\, M \left(\frac1 2-\frac{b }{2 } -\frac{i k}{\alpha}, 1- \frac{2 i
k}{\alpha}; y\right)\,, \ \label{MatsumotoWF} \eea with $C$  a real normalization constant, and \bea \tilde
A=\frac{\Gamma(-\frac{2 i k}{\alpha})} {\Gamma\left(\frac1 2- \frac{b }{2}- \frac{ik}{\alpha}\right)}\ .\label{A} \eea

Let us now consider the bound states. Their wave functions  are given in \cite{FRW}  as \ben \psi(x)\propto \exp(-\kappa
e^{-\al x}/\al)e^{-\al (\nu-b /2  +1/2)x} L^{b -2\nu-1}_{\nu}(2\kappa e^{-\al x}/\al)\ ,\label{Morse_bound_pre}
\een
 where
  \ben L^\lambda_n(x)=\sum_{m=0}^{n}(-1)^m {{n+\lambda} \choose
{n-m}}\frac{x^m}{m!}\label{Laguerre_assoc} \een
 are the associated Laguerre polynomials $L^\lambda_n(x)$. The energies are
 \ben
 E_\nu=-\frac{\hbar^2\al^2}{2m}(\nu-b /2 +1/2)^2 ,\label{energy_Morse}
 \een
 for integer $\nu\in [0,\lfloor b /2 \rfloor]$, where $\lfloor a\rfloor$ is the smallest integer less than $a$.

%%%%%%%%%%%%%%%%%%%%%%%%%%%%%%%%%%%%%%%%%%%%%%%%%%%%%%%%%%%%%%%%%%%%%%%%
\vskip.25cm \noindent{\bf Acknowledgements}\hfill\break This research was supported in part by a Discovery Grant from the
Natural Sciences and Engineering Research Council of Canada and by the School of Graduate Studies of the University of
Lethbridge. \hfb M.W. thanks the Instituto de Matem\'aticas de UNAM in Morelia, M\'exico, for its warm hospitality. He is
grateful to S. Kryukov for showing him the relevance of Mellin transforms in solving $\ast$-eigenvalue equations. We also thank
W. Chemissany, S. Das, A. Dasgupta, S. Neale and S. Sur for comments; and R. Badhuri for a useful e-mail message regarding
supersymmetric quantum mechanics.

%%%%%%%%%%%%%%%%%%%%%%%%%%%%%%%%%%%%%%%%%%%%%%%%%%%%%%%%%%%%%%%%%%%%%%%%

\end{document}